\documentclass[prl,twocolumn,superscriptaddress,preprintnumbers,amsmath,amssymb]{revtex4}
\usepackage{graphicx}
\usepackage{amsmath}
\usepackage{amssymb}
\usepackage{epsfig}
\usepackage{wasysym}
\usepackage{bbm}
\renewcommand{\narrowtext}{\begin{multicols}{2} \global\columnwidth20.5pc}
\renewcommand{\v}[1]{{\bf #1}}

\def\be{\begin{eqnarray}}
\def\ee{\end{eqnarray}}

\newcommand{\Eq}[1]{Eq.~(\ref{#1})}
\newcommand{\Fig}[1]{Fig.~(\ref{#1})}

\newcommand{\nd}{{\vphantom{\dagger}}}

\begin{document}

\title{Funtional Renormalization Group Study of the Pairing
Symmetry and Pairing Mechanism of the FeAs Based High Temperature Superconductors}

\author{Fa Wang}
\author{Hui Zhai}
\author{Ying Ran}
\author{Ashvin Vishwanath}
\author{Dung-Hai Lee}

\affiliation{
Department of Physics,University of California at Berkeley,
Berkeley, CA 94720, USA}
\affiliation{Materials Sciences Division,
Lawrence Berkeley National Laboratory, Berkeley, CA 94720, USA}

\begin{abstract}
We apply the fermion functional renormalization group method to determine the pairing symmetry and pairing mechanism of the  FeAs-Based materials. Within a five band model with pure repulsive interactions, we find an electronic-driven superconducting pairing instability. For the doping and interaction parameters we have examined, extended $s$-wave, whose order parameter takes on opposite sign on the electron and hole pockets, is always the most favorable pairing symmetry. The pairing mechanism is the inter Fermi surface Josephson scattering generated by the antiferromagnetic correlation.
\end{abstract}

\date{\today}
\maketitle

The copper oxide high temperature superconductors (the cuprates) have stimulated a significant part of condensed matter physics development in the past twenty years. Two years ago, a new non-copper superconducting compound, the iron pnictides, was discovered\cite{Tc-Japan}. In the last few months, by various element substitution, the superconducting transition temperature of this new class of superconductor has been raised to 55K\cite{Tc-2,Tc}. This has stimulated a flurry of interests in their material and physical properties. Examples of iron pnictides include LnFeAsO$_{1-x}$F$_x$\cite{Tc-Japan,Tc-2,Tc},  La$_{1-x}$Sr$_x$FeAsO\cite{haihu}, and Ba$_{1-x}$K$_x$Fe$_2$As$_2$\cite{BFEAS}.

Unlike the cuprates, the stoichiometric parent compounds of the iron pnicides are  {\it metallic} (or semi-metallic)\cite{nanlin} antiferromagnets\cite{neutron} rather than antiferromagnetic Mott insulators. For example, recent angle-resolved photoemission spectroscopy on single crystal BaFe$_2$As$_2$ has revealed residual Fermi surface in the antiferromagnetic state\cite{donglai}. On the other hand, similar to the cuprates, the antiferromagnetic order is quickly replaced by superconductivity as a function of doping\cite{Competing}, suggesting an intimate relation between antiferromagnetism and superconductivity. Currently it is generally felt that progresses in understanding this new class of high T$_c$ superconductor will also have impacts on the two decade old problem - the  cuprates.

If one reviews the physics of the cuprates, it is easy to spot the prominent role played by the pairing symmetry. At the present time early experiments on the iron pnictides have produced
conflicting claims on this issue\cite{linenode,chien}. As far as theoretical proposals are concerned, pairing symmetries spread all over the map \cite{sd,LDA-2}.

The purpose of this paper is to determine the pairing symmetry and pairing mechanism theoretically.
However, because of the intermediate coupling nature of these material \cite{kotliar}, it is not obvious where to start.
For example, if one takes a model of the bandstructure and adds repulsive local interaction to it, it is easy to show that at the mean-field level there is no superconducting instability in reasonable parameter range. In addition, unlike the cuprates, there is no strong coupling (the $t^2/U$ ) expansion that can lead to antiferromagnetic exchange which in turn can produce pairing (at least at the mean-field level). In the process of searching for a suitable calculational scheme we come across the functional renormalization group (FRG) method\cite{RG}. The numerical version of this  method was implemented in Ref.~\cite{schulz,Honerkamp} in the study of the cuprates. Interestingly it successfully reproduced both the antiferromagnetic and the d-wave pairing tendencies.
The advantage of this method lies in its ability to generate effective interactions which are otherwise absent at the bare level. In addition, by monitoring the growth of the renormalized interaction, it is possible to pin down the cause of different types of order. Finally in view of the fact that the electron-electron correlation of the iron pnictides is weaker than that of cuprates\cite{kotliar}, we felt that the FRG method will have a good chance to succeed.

To model the band structure, we take the most complete tight binding model we can find\cite{kuroki}, then we add the Hubbard-like and Hunds-like local interaction to it. Like most works in the literature we focus on the As-Fe-As tri-layers, and view the arsenics as mediating hopping between the five iron 3d orbitals. The resulting 
tight-binding Hamiltonian reads
\be
 \hat{H}_0=\sum_{\v{k},s}\sum_{a,b=1}^5 c_{a\v{k} s}^\dagger K_{ab}(\v{k})c_{b\v{k} s}^\nd
\label{h0}\ee
where $c_{a\v{k} s}$ annihilates a spin $s$ electron in orbital $a$ and momentum $\v{k}$.The parameters used in constructing  $K_{ab}(\v k)$ can be found in Ref.~\cite{kuroki}. However, in our calculation a gauge has been chosen so that all elements of $K_{ab}(\v k)$ are real.
Doping is controlled by adding a chemical potential term to \Eq{h0}. For sufficiently large electron doping,
there are two different hole pockets centered at $\v k=(0,0)$ and two electron pockets centered at $(\pi,0)$ and $(0,\pi)$.
For hole doping, or small electron doping, an extra hole pocket is present at $\v k=(\pi,\pi)$.
These features are consistent with other DFT calculations\cite{LDA,LDA-2}.

We describe the electron-electron interaction by
\begin{widetext}
\be
\hat{H}_{\rm int}=\sum_{i}\Big\{U_1\sum_{\alpha}
 n_{i,\alpha,\uparrow}n_{i,\alpha,\downarrow}+U_2\sum_{\alpha< \beta}n_{i,\alpha} n_{i,\beta}+
 J_H\Big[\sum_{\alpha< \beta}\sum_{s,s'}c_{i\alpha s}^\dagger c_{i\beta s'}^\dagger c_{i\alpha s'}^\nd c_{i\beta s}+ (c^\dagger_{i\alpha\uparrow}c^\dagger_{i\alpha\downarrow}c_{i\beta\downarrow}c_{i\beta\uparrow}+h.c.) \Big]\Big\}.
 \ee
 \end{widetext}
Here $i$ labels the sites of a square lattice, $s,s'=\uparrow,\downarrow$, and $n_{i,\alpha}=n_{i,\alpha,\uparrow}+n_{i,\alpha,\downarrow}$ is
the number operator associated with orbital $\alpha$. This Hamiltonian includes the intra and inter orbital Hubbard $U_1$ and $U_2$,
the Hund's interaction $J_H$, and the inter orbital pair hopping.
The total Hamiltonian
$\hat{H}=\hat{H}_0+\hat{H}_{\rm int}$ is the starting point of our study.
The bare interaction parameters we use throughout the rest of the paper are $U_1=4.0,\,U_2=2.0,$ and $J_H=0.7$ eV.
The energy scale $4$eV is taken from Ref.~\cite{kotliar}, and we have chosen the parameters so that it approximately satisfies the relation $U_1=U_2+2J_H$ \cite{relation}.

We have checked that mean field theory done on the bare Hamiltonian, $\hat{H}_0+\hat{H}_{\rm int}$, has
{\em no} superconducting instability for realistic interaction parameters $U_{1,2}>J_H$.
In following we shall show that as the high energy electronic excitations are recursively integrated out an effective interaction that drives extended s-wave pairing is generated.

\begin{figure}[b]
\includegraphics[scale=.45]{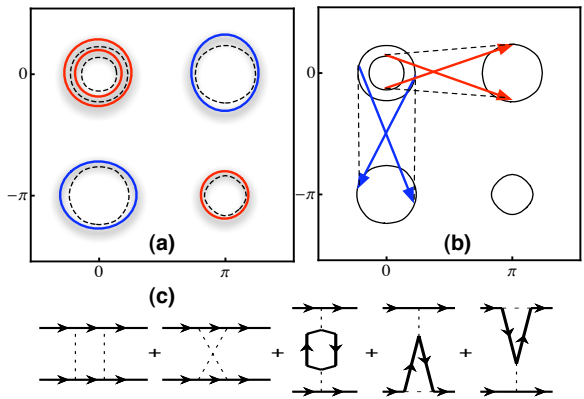}
\caption{(a) A schematic representation of the gap functions on different Fermi surfaces. The Fermi surfaces are shown in black dashed lines. The width of the  region between the solid line and dashed line indicates the magnitude of the gap function, and the sign is represented by the colors. (b) Two (red and blue arrows) typical inter Fermi surface pair transfer (Josephson) processes that drive pairing. (c) The five Feynman diagrams contributing to the renormalization of the vertex function.}\label{NRG}
\end{figure}

Many of the technical details of our FRG can be found in Ref.~\cite{Honerkamp}. At the end of the present paper we discuss few technical challenges in generalizing the method to the present multi-band and multi-Fermi surface situation.
In brief, we divide the Brillouin zone (BZ) into $N$ patches, and at each renormalization iteration we sum over the five one-loop Feynman diagrams in \Fig{NRG}(c) in computing the renormalized 4-point vertex function
$V(\v k_1,a;\v k_2,b;\v k_3,c;\v k_4,d)$.  Here $a,b,c,d=0,\dots,4$ labels
the five different bands. The momenta $\v k_1,..,\v k_3$ are on the Fermi surfaces. (Due to momentum conservation the fourth momentum $\v k_4$ is determined.) The convention is chosen so that the spin associated with $\v k_3$ ($\v k_4$) is the same as that associated with $\v k_1$ ($\v k_2$). Like in Ref.~\cite{Honerkamp}, approximations (such as ignoring the frequency dependence of the vertex function, and projecting the external momenta onto the fermi surfaces) are made in our calculation due to practical limitations.

From the renormalized vertex function we extract the effective interaction in the Cooper channel as follows
\be
&&V^{sc}_{s,t}(\v k,a;\v p,b) \nonumber \\ &&=V(\v k,a;-\v k,a;\v p,b;-\v p,b)\pm V(-\v k,a;\v k,a;\v p,b;-\v p,b).\nonumber
\ee
Here the upper/lower sign is for singlet/triplet pairing, respectively.
After the BZ discretization  $\v k$ and $\v p$ only take on a finite number of values. Therefore we can treat $V^{SC}_{s,t}$ as matrices with $(\v k,a)$ and $(\v p,b)$ as indices.
The eigenvalues of this matrix are the effective interaction strength in each pairing channel and the eigenvectors are the pairing form factors.

\begin{figure}
\includegraphics[scale=0.68]{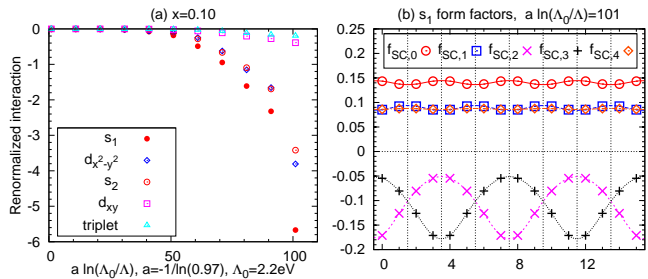}
\caption{(a) The $N=32$ RG evolution of the scattering amplitude in the top four attractive Cooper channels for $0.1$ electron doped system(6.1 electrons per site). Here $\Lambda_0=2.2 eV$ is the initial energy cutoff, and $\Lambda$ is the running energy cutoff. The constant $a=-1/\ln(0.97)$. (b) The gap function associated with the most attractive pairing channel determined from the
 final ($a\ln(\Lambda_0/\Lambda)=101$) renormalized interaction vertex function. It has a full gap and  exhibits significant amplitude modulation on the electron pockets (Fermi surface 2 and 3). More importantly, the sign of the gap function is different on the electron and hole packets.}\label{flow1}
\end{figure}

\begin{figure}
\includegraphics[scale=0.5]{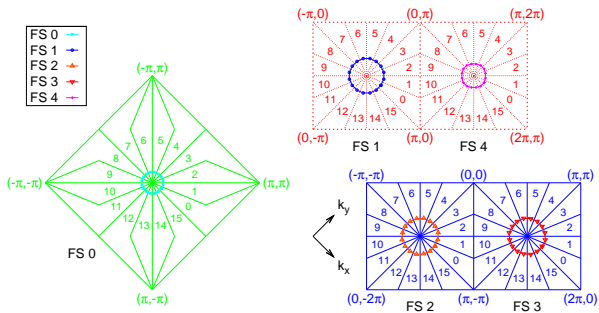}
\caption{The Brillouin zone patching scheme. Here the five disjoint Fermi surfaces are labeled from 0 to 4.
The left panel is for the band with smaller hole pocket at $(0,0)$ (FS 0).
The right top panel is for the band with larger hole pocket at $(0,0)$ (FS 1) and hole pocket at $(\pi,\pi)$ (FS 4).
The right bottom panel is for the electron pockets (FS 2 and FS 3).
}\label{fig:patching}
\end{figure}

{\bf The results for $10\%$ electron doping}: The five disjoint Fermi surfaces at this doping level are schematically illustrated in \Fig{NRG}(b). For $N=32$ (the number of discretized BZ patches, see \Fig{fig:patching}) the few lowest eigenvalues of $V^{SC}_{s,t}$
as a function of the RG running cutoff $\ln(\Lambda_0/\Lambda)$ is shown in \Fig{flow1}(a).
Here $\Lambda_0,\Lambda$ are the initial and the running energy cutoffs.
As $\ln(\Lambda_0/\Lambda)$ increases the most attractive pairing channels has extended $s$-wave symmetry $s_1$. In particular, the sign of superconducting order parameter is opposite on the electron and hole pockets.
The associated gap function, $f_a(\v k)$ (here $a$ labels the Fermi surfaces),
is shown in \Fig{flow1}(b), and  \Fig{NRG}(a) is a schematic representation of it.
Interestingly, $|f_a(\v k)|$ shows substantial
variation on the electron pockets (2 and 3).
This variation can be understood qualitatively by considering the following form factor $A_1+A_2 (\cos k_x+\cos k_y)$. For example, if we expand $(k_x,k_y)$ around $(\pi,0)$, via $k_x=q_x+\pi$, $k_y=q_y$, the above form factor becomes $A_1 - A_2 (\cos q_x-\cos q_y)$. As a result, it contains a d-wave like component as $\v q$ goes around the electron pockets. If $A_2$ is sufficiently big such form factor can exhibit nodes on the electron pocket despite the fact it has s-wave symmetry. Although for the parameters we have studied the gap function has no nodes on any Fermi surfaces, we can not rule out the possibility that for other parameter choice the s-wave gap function can have nodes on the electron pockets.
In \Fig{flow1}(a) we also show the RG flow of the three leading sub-dominant pairing channels $d_{x^2-y^2}$, extended-$s$ with nodes $s_2$, and $d_{xy}$. For the parameter and doping we have studied, triplet pairing is not favored.

Most significantly, from our calculation the pairing mechanism can be determined. By monitoring the RG evolution of the vertex function, the following sequence of events are observed. First of all, the bare Hamiltonian  contains interaction that can drive SDW. Upon RG these scattering vertices grow stronger. Among these growing scattering amplitudes there are ones that also cause inter-pocket pair scattering as shown in \Fig{NRG}(b). They serve as the ``seed'' for the growth of other inter-pocket pair scattering processes in subsequent RG steps. It is important to note that the inter-pocket pair scattering drives pairing even when it is positive\cite{tband}. In that case the superconducting order parameter will have opposite sign on the two Fermi pockets, as in our extended s-wave pairing case discussed above.
From these observations we conclude that pairing is driven by the antiferromagnetic correlation.

\begin{figure}
\includegraphics[scale=0.68]{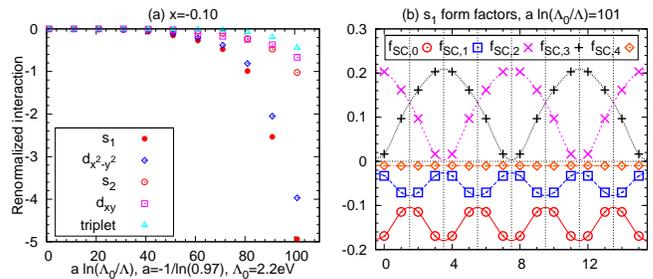}
\caption{(a) The $N=32$ RG evolution of the scattering amplitude in the top four attractive Cooper channels for $0.1$ hole doped system(5.9 electrons per site), where $\Lambda_0=2.2 eV$ and $a=-1/\ln(0.97)$. (b) The gap function associated with the most attractive (s-wave) pairing channel determined from the
 final ($a\ln(\Lambda_0/\Lambda)=101$) renormalized interaction vertex function. It has a full gap, and exhibits significant amplitude modulation on Fermi surfaces 0, 1, 2 and 3.}\label{flow2}
\end{figure}

{\bf The results for $10\%$ hole doping}:
The Fermi surface topology for this doping value is the same as that
of $10\%$ electron doping (\Fig{NRG}(a)).
The few lowest eigenvalues of the $V^{SC}_{s,t}$ matrices as a function
of the RG running cutoff is shown in \Fig{flow2}(a). Again the
most attractive pairing channel is extended $s$-wave like. The
gap function associated with it is
shown in \Fig{flow2}(b), and similarly represented by \Fig{NRG}(a).
Compared to \Fig{flow1}(b) the gap functions on the $(\pi,\pi)$ hole packets is
considerably smaller. The mechanism for the electron doping applies to this case as well.

Thus for both electron and hole doped cases discussed above extended $s$-wave is
the pairing symmetry. In contrast, non-s-wave symmetry such as d-wave or triplet pairing are not favored.
Based on the above results, {\em we predict that the extended $s$-wave is the
pairing symmetry of the FeAs based superconductors.}

We have also tried several other sets of interaction parameters, for instance, varying $J_H$ over the range $0\leq J_H<U_{1,2}$, varying the ratio and the overall magnitudes of $U_{1,2}$. In addition we have varied the doping while maintaining the topology of hole Fermi pockets at $\Gamma$ and the  electron pockets at $M$. The result that the gap function assumes s-wave symmetry and takes opposite phase on the electron and hole pockets remains unchanged. This is even so for significant electron doping so that the hole pocket at $(\pi,\pi)$ disappears. The qualitative nature of our results is not affected by the choice of the initial cutoff, as long as it is comparable to the band width. Interestingly, the degree of gap variation on the electron and hole pockets does depend on interaction parameters and doping. For relatively weak interaction parameters, we have tried $U_1=2.4eV$, $U_2=1.2eV$ and $J_{H}=0.42eV$, the results remains qualitatively similar. However, for $U_1=1.2eV$, $U_2=0.9eV$ and $J_{H}=0.15eV$, we have not observed any divergence of any scattering vertices to the lowest temperature ($0.1 meV$) we have studied.

{\bf FRG details}:
The essential complication of our FRG in comparison with that in Ref.~\cite{Honerkamp} is the fact that the Fermi surface(s) (if any) associated with each band is different. This requires us to discretize the BZ differently for different bands (see \Fig{fig:patching}).
For the band with electron pockets, the BZ is chosen to be two $45^{\circ}$-rotated squares, centered at $(0,-\pi)$ and $(\pi,0)$ (see the bottom panels of \Fig{fig:patching}).
Each square is divided into $N/2=16$ patches by radial lines from its center.
The BZ discretization for the band containing the large hole pocket at $(0,0)$ and $(\pi,\pi)$ is similar, but with the two $45^{\circ}$-rotated squares centered at $(0,0)$ and $(\pi,\pi)$. The BZ discretization of the band containing the small hole packet at $(0,0)$ is the same as what is used in Ref.~\cite{Honerkamp}.

Like in Ref.~\cite{Honerkamp} the value of the renormalized vertex function at a given $(\v k_1,\v k_2,\v k_3)$ in the remaining (un-integrated) BZ region is made equal to the value of such function with  $\v k_1,\v k_2,\v k_3$ projected to the closest Fermi surface segments. As a result, for the three bands with five disjoint Fermi surfaces each  discretized into $M$ segments (in our case $M=16$), we need to recompute $1.536\times 10^6$ interaction vertices at each step of the RG. The top and bottom bands that have no Fermi surface are ignored.
 Due to the existence of many degenerate points in the bandstructure, the gauge choice of the numerical Bloch wavefunctions needs to be fixed with caution.

{\bf Conclusion} We have performed a numerical functional renormalization group calculation to determine the pairing symmetry of the FeAs superconductors. We find $s$-wave with the opposite sign on the electron and hole pockets is the pairing symmetry for both electron and hole dopings.
By monitoring the growth of the renormalized vertex function, we find the pairing mechanism is driven by the antiferromagnetic correlations. We hope this prediction will be scrutinized by future experiments.

Acknowledgement: We thank Henry Fu and C. Honerkamp for most helpful communications.
DHL was supported by DOE grant number DE-AC02-05CH11231. AV was supported by LBNL DOE-504108.

\end{document}